\documentclass[iop,apjl]{emulateapj}
\usepackage{graphicx}
\usepackage{amsmath}
\usepackage{amsfonts}
\usepackage{amssymb}

\def\Ss/{\mbox{$S_{*}$}}
\def\Sg/{\mbox{$S_{gal}$}}

\def\obj/{Balbinot 1}
\def\fsat/{\texttt{FindSat}}

\shorttitle{A new Milky Way halo star cluster in the Southern Galactic Sky}
\shortauthors{Balbinot et al.}

\begin{document}

\title{A new Milky Way halo star cluster in the Southern Galactic Sky}

\author{E. Balbinot\altaffilmark{1,2}, B. X. Santiago\altaffilmark{1,2}, L. da
    Costa\altaffilmark{2,3}, M. A. G. Maia\altaffilmark{2,3}, S. R.
    Majewski\altaffilmark{4}, D. Nidever\altaffilmark{5}, H. J.
    Rocha-Pinto\altaffilmark{2,6}, D. Thomas\altaffilmark{7}, R.H.
    Wechsler\altaffilmark{8}, B. Yanny\altaffilmark{9}}

\email{Correspondece to: balbinot@if.ufrgs.br}

\altaffiltext{1}{Instituto de F\'isica, UFRGS, CP 15051, Porto Alegre, RS 91501-970, Brazil}
\altaffiltext{2}{Laborat\'orio Interinstitucional de e-Astronomia - LIneA, Rua Gal. Jos\'e Cristino 77, Rio de Janeiro, RJ - 20921-400, Brazil}
\altaffiltext{3}{Observat\'orio Nacional, Rua Gal. Jos\'e Cristino 77, Rio de Janeiro, RJ - 22460-040, Brazil}
\altaffiltext{4}{Department of Astronomy, University of Virginia, Charlottesville, VA 22904-4325, USA}
\altaffiltext{5}{Department of Astronomy, University of Michigan, Ann Arbor, MI 48109-1042, USA}
\altaffiltext{6}{Observat\'orio do Valongo, Universidade Federal do Rio de Janeiro, Rio de Janeiro RJ 20080-090, Brazil}
\altaffiltext{7}{Institute of Cosmology and Gravitation, University of Portsmouth, Portsmouth, Hampshire PO1 2UP, UK}
\altaffiltext{8}{Kavli Institute for Particle Astrophysics and Cosmology, SLAC
National Accelerator Laboratory; Department of Physics, Stanford University, Stanford, CA 94305, USA}
\altaffiltext{9}{Fermi National Laboratory, P.O. Box 500, Batavia, IL 60510-5011, USA}

\begin{abstract} 
We report on the discovery of a new Milky Way companion stellar system located
at $(\alpha_{J2000},\delta_{J2000}) = (22^h10^m43.15^s, 14^\circ56\arcmin
58.8\arcsec)$. The discovery was made using the eighth data release of SDSS after
applying an automated method to search for overdensities in the Baryon
Oscillation Spectroscopic Survey footprint. Follow-up observations were
performed using CFHT/MegaCam, which reveal that this system is comprised of an
old stellar population, located at a distance of $31.9^{+1.0}_{-1.6}$ kpc, with
a half-light radius of $r_h= 7.24^{+1.94}_{-1.29}$ pc and a concentration parameter of
$c = \log_{10}(r_t/r_c) = 1.55$. A systematic isochrone fit to its
color-magnitude diagram resulted in $\log(age/yr) = 10.07^{+0.05}_{-0.03}$ and
$[Fe/H] = -1.58^{+0.08}_{-0.13}$. These quantities are typical of globular
clusters in the MW halo. The newly found object is of low stellar mass, whose
observed excess relative to the background is caused by $95\pm6$ stars. The
direct integration of its background decontaminated luminosity function leads
to an absolute magnitude of $M_V = -1.21\pm0.66$. The resulting surface
brightness is $\mu_V = 25.90$ mag/arcsec$^2$. Its position in the $M_V$ vs.
$r_h$ diagram lies close to AM4 and Koposov 1, which are identified as star
clusters.  The object is most likely a very faint star cluster --- one of the
faintest and lowest mass systems yet identified.
\end{abstract}


\section{Introduction}

Recent large surveys such as Sloan Digital Sky Survey (SDSS) and the 2-Micron
All Sky Survey (2MASS) have delivered an enormous amount of data about the
stellar populations of the Milky Way (MW).  These studies have probed a new
regime in parameter space of Milky Way satellites, by significantly expanding
the volume over which the faintest systems can be detected, and have revealed a
wealth of new objects.  One striking discovery from these surveys is the myriad
of substructures that populate the MW structural components, including stellar
streams from disrupting satellite galaxies and tidal tails from globular
clusters \citep[e.g][]{rock, majewski03, rocha04, newberg10}.  It has included
the exciting discovery of a new class of faint dwarf galaxies
\citep[e.g.][]{willman05, belokurov06, irwin07, walsh}, which may elucidate our
understanding of small-scale structure in the dark matter distribution and of
galaxy formation at the lowest masses. In addition, a handful of very faint
stellar systems have been identified in the outer halo \citep{koposov07,
belokurov10, munoz, fadely}, with somewhat different properties from more
massive clusters.

Studies of extragalactic star cluster systems have also revealed the existence
of diffuse and low surface brightness stellar systems around luminous galaxies,
such as Faint Fuzzies (FF) and Diffuse Star Clusters (DSC) \citep{larsen00,
peng06}.  Studying nearby counterparts of these elusive objects will allow us to
better constrain their structure, dynamics, and formation histories.

A significant population of MW satellites is very likely still left undiscovered
\citep{tollerud2008, willman2010}.  Future deep surveys, such as the Dark Energy
Survey\footnote{http://darkenergysurvey.org} and the Large Synoptic Survey
Telescope \footnote{http://www.lsst.org}, will allow very faint systems to be
probed much further out in the Galactic Halo, and are likely to provide
important new clues to the formation mechanism of globular clusters, to the
hierarchical buildup of the Milky Way system, to galaxy formation physics in low
masses systems, and to the abundance and properties of the smallest scale
structures in the Universe \citep{rossetto2011}.

In this work we present the discovery of a new stellar system in the MW halo.
The object is located at $(\alpha_{J2000},\delta_{J2000}) = (22^h10^m43.15^s,
14^\circ56\arcmin 58.8\arcsec)$, or $(l,b) = (75.1735^o, -32.6432^o)$, in the
Pegasus constellation. Its standard SDSS object name is SDSS J2211+1457,
although in this work we choose to call it by a shorter name, \obj/. This
paper is organized as follows. In section 2 we describe the data and
methods that led to this discovery. In section 3 the follow-up observations and
data reduction are briefly explained. In section 4 we quantify the properties
of the object in section 5 and discuss the nature of \obj/ in comparison to
other satellite systems discovered in the MW. This object is likely a very old
stellar cluster, one of the lowest mass and lowest surface brightness objects
yet detected.

\section{Data and Method}

The Baryon Oscillation Spectroscopic Survey (BOSS) is one of the four
surveys that belong to the Sloan Digital Sky Survey III (SDSS-III).
Its imaging stage is now complete with photometry in the $ugriz$
system for $\sim 8 \times 10^7$ sources covering a region of $\sim
2000$ deg$^2$ in the southern Galactic hemisphere was released in
January 2011 as part of the SDSS Data Release 8 \citep[DR8;][]{aihara2011}. 
So far, no systematic search for MW faint satellites has been published
in the BOSS footprint. Our discovery data were taken from the DR8 
\texttt{PhotoPrimary} view table. A {\it clean} sample was defined by selecting 
those objects flagged as reliable stars or galaxies (see detection process 
below).

To search for stellar systems in a large area of the sky, we developed an
automated tool called \fsat/. The algorithm closely follows the work of
\citet{koposov08} and \citet{walsh}.  There are three main steps in the
algorithm. First a color-magnitude cut based on stellar evolutionary models is
applied to enhance the presence of old metal-poor stellar populations relative
to field stars. Next, \fsat/ uses this filtered stellar sample and creates a
density map on the sky plane.  Finally, the density map is convolved with a
kernel that is the difference between two Gaussians, one with the angular size
of a typical Milky Way (MW) satellite ($\sim 4-8 \arcmin$), and the other much
wider, used to smooth out any remaining large scale structure on the map.

Source detection is performed on the convolved map using \texttt{SExtractor}
\citep{bertin}. To circumvent misidentification due to poor star-galaxy
separation, we apply identical steps to build density maps of sources
classified as galaxies by the SDSS reduction pipeline and apply SExtractor to
them.  For a given detected source, \fsat/ then compares the signals in
the smoothed stellar and galaxy density images as determined by SExtractor,
(\Ss/ and \Sg/, respectively).  Known faint satellites tend to fall in the low
\Sg/ and high \Ss/ locus of \Sg/  vs. \Ss/ plane. We use this region to select
our new satellite candidates. We also compare the position of these new
overdensities with those of known objects found in Abell, Globular Clusters,
and NGC catalogs, to remove any known object that may share the same locus of
the \Sg/  vs. \Ss/ plane.

For the BOSS region we ran \fsat/ using color-magnitude cuts based on four
Padova evolutionary models \citep{girardi} with ages of 8 and 14 Gyr and $Z =
[0.001, 0.006]$. We tuned our CMD filter to search search for objects at five
distance moduli $(m-M)=[18.0, 19.0, 20.0, 21.0, 22.0]$.  Visual inspection of
the detected candidates was carried out, revealing \obj/ as a faint overdensity
of blue sources consistent with them being stars. However, deeper and higher
spatial resolution images were required to investigate the nature of \obj/.

\section{Follow-up observations}

In the first semester of 2012, we obtained follow-up observations of \obj/
using MegaCam, which is installed in the Canada-France-Hawaii-Telescope (CFHT).
MegaCam is a mosaic 36 CCDs, each with $2048\times4612pix$. The total field of
view (FOV) is $0.96\times0.94deg$, with a pixel scale of $\sim$ 0.19
$\arcsec/pix$.

We designed the observations to reach the magnitude of the main sequence
turn-off (MSTO) of a $10$ Gyr stellar population at $150$ kpc distance with
$S/N \sim 10$. Exposure times of $2800$ s in the $g$ band and $3800$ s in the $r$
band were required. The total exposure time was divided into 6 dithering
exposures for each passband to avoid scattered light from bright stars,
saturation, and also to cover the smaller gaps between the CCDs. The
observations were carried out under photometric conditions, seeing was always
below $0.8\arcsec$, and the airmass below $1.15$.

\begin{figure} \begin{center}
    \includegraphics[width=0.5\textwidth]{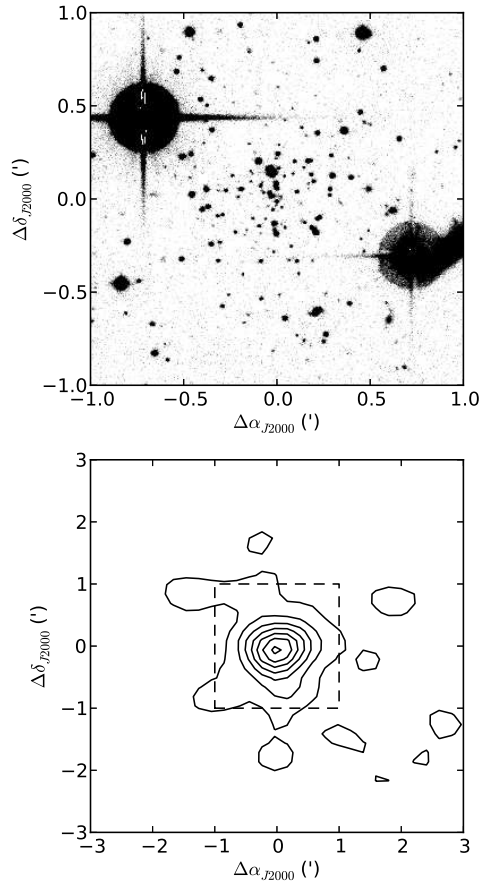}
    \caption{\textit{Bottom panel:} the isodensity contours built using the list
    of sources detected by Terapix. The contour levels are for $[8, 16, 24, 32,
    40]$ stars$/arcmin^2$.  The dashed box shows a $1\arcmin\times1\arcmin$ region
    for which we show the combined $g$ band image in the \textit{top panel}.}
\label{CFHT_g} 
\end{center}
\end{figure}

The basic reduction of the images (overscan and bias subtraction, and
flatfielding) was done by the CFHT team. The images were geometrically
corrected, registered and co-added by Terapix, which uses the Astromatic
toolkit\footnote{http://www.astromatic.net/}. Therefore, source detection is
based on SExtractor. Point spread function (PSF) photometry was carried out
using \texttt{DAOPHOT} \citep{stetson}.  The input coordinates list for the
stars measured using \texttt{DAOPHOT} was the one provided by Terapix. This list
presented less spurious detections than the one from \texttt{DAOFIND}.
Photometric calibration was achieved using bright and non-saturated ($19 <
g_{SDSS} < 21$) stars in the observed field, whose instrumental magnitudes were
then compared to those from SDSS.  We found 536 matches which we used to find a
calibration equation composed of a photometric zero point and a color term. The
equation coefficients were found by means of a least square fit using sigma
clipping rejection. Our mean calibration residuals are of $0.024$ in the $g$
band and $0.019$ in the $r$ band.

The observed MegaCam field shows little differential reddening. Using the dust
maps from \citet{schlegel} we find a value of E(B-V) = 0.060 at position of
Balbinot 1 and a standard deviation of 0.0014 across the whole field. 

Thus, differential reddening is not likely to contribute significantly to our
analysis. Nonetheless we choose to correct for reddening using the value of
E(B-V) at the position of each star in the MegaCam field. We adopt $R_V = 3.1$
and the coefficients from \citet{cardelli} to compute the extinction in the CFHT
passbands. 

Figure \ref{CFHT_g} shows a stacked $g$ band image centered on \obj/ (top
panel). We also show  isodensity contours of our detected sources (bottom
panel). A clear overdensity of stars is seen on both panels of the figure very
close to where the \fsat/ candidate was originally identified.

\section{Results}

\begin{figure}
   \begin{center}
   \includegraphics[width=0.5\textwidth]{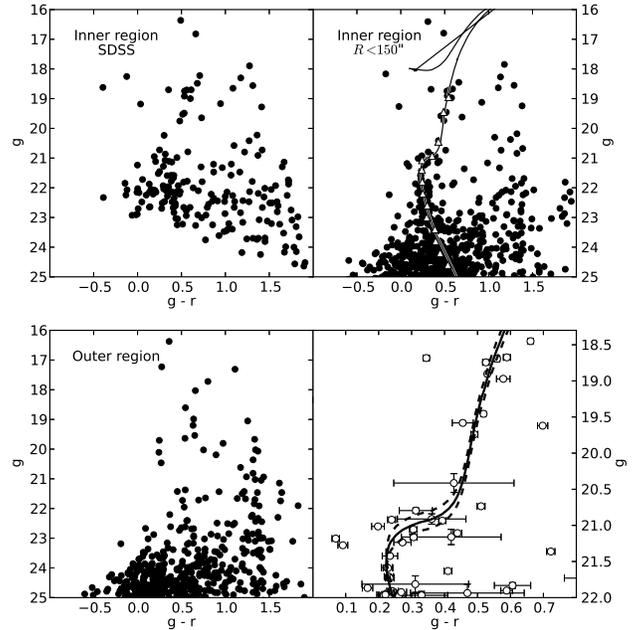}
   \caption{\textit{Top left panel}: $(g-r) \times g$ CMD for SDSS measured
       sources inside a radius of $150\arcsec$ from the visual center of \obj/.
       \textit{Top right panel}: the CMD for the same region as the previous
       panel but showing sources from our CFHT photometry. The white triangles
       show the object ridge line. \textit{Bottom left panel:} CMD for sources
       located in a ring far away from the object center. The area of the ring
       is the same as the area of the inner region. In the \textit{bottom right
       panel} we show zoomed in version of the top right panel around the MSTO
       region; we also show the photometric errors for each source. The best fit
       isochrone is shown as a solid line.  The isochrones with upper and lower
   limits of distance modulus are shown by the dashed lines.} 
   \label{CMD_comp}
   \end{center}
\end{figure}

From Figure \ref{CFHT_g} it is evident that there is a concentration of
stellar-like objects in the observed region. To confirm its stellar nature we
must analyze its CMD. In Figure \ref{CMD_comp} we show the extinction corrected
$(g-r) \times g$ CMDs. In the top left panel we show the CMD based on the SDSS
discovery data. On the top right panel we show the CMD from our follow up CFHT
images. In both cases the stars are restricted to a radius of $150\arcsec$ from
the visual center of \obj/. To discard \obj/ as a density fluctuation of field
stars we also show (\textit{bottom left panel}) the CMD for a ring well away
from the object center covering the same area on the sky.  By comparing these
panels we not only clearly confirm the excess of stars around the position of
\obj/ but also conclude that the distribution of these stars on the CMD plane is
consistent with a simple stellar population (SSP). An obvious MSTO is seen at
$(g-r) \simeq 0.21$ and $g \simeq 21.4$, in connection to an RGB stretching up
to $(g-r) \simeq 0.6$ and $g \simeq 18.5$.

The bottom right panel in Figure \ref{CMD_comp} shows the MSTO region in detail.
We also show the best fitting Padova isochrone \citep{girardi}. It corresponds
to an age of $11.7^{+1.4}_{-0.8}$ Gyr and an abundance of $Z=0.0005$, or [Fe/H]
$= -1.58^{+0.08}_{-0.13}$ at a distance of $d_{\odot} = 31.9^{+1.0}_{-1.6}$ kpc.
In the same panel we show the effect of varying the distance modulus within its
fit uncertainties. The isochrone fit was performed by means of a $\chi^2$
minimization of the CMD distance between the cluster ridge line and the
isochrone set.  Our model grid goes from 9.8 to 10.12 in $log(age/yr)$ with a
step of 0.01, and from 0.0001 to 0.004 in Z, with steps of 0.0001. We adopt a
step of 0.01 in distance modulus and a range from 16.0 to 18.0. 
The best model is the one that minimizes $\chi^2$. The uncertainties are
derived from models that have values of $\chi^2 = \chi^2_{min} + 1$.

We stress that the best fit isochrone and its associated parameters are somewhat
dependent on the stellar evolution model and on the model grid. In fact, our
quoted metallicity uncertainty is comparable to the Padova grid resolution for
this parameter. Also, the discrepancies among different evolutionary models are
much larger than our quoted uncertainties. For a comprehensive approach to this
problem we refer to \citet{kerber}.

\subsection{Radial profile and half-mass radius}

\begin{figure}
   \begin{center}
   \includegraphics[width=0.5\textwidth]{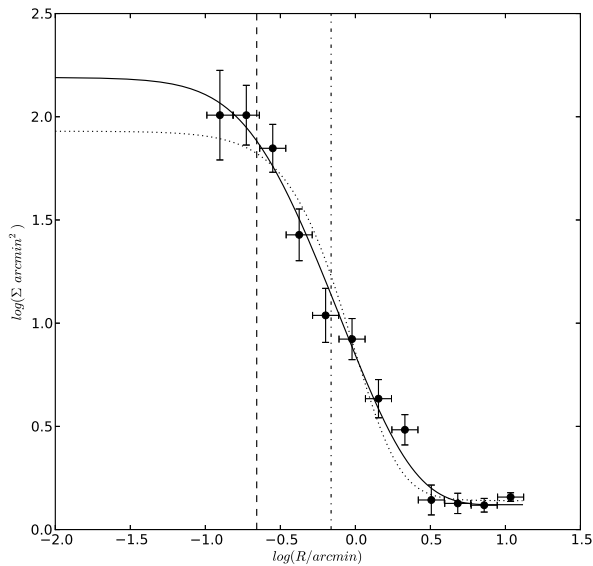}
   \caption{The radial density profile for \obj/ with $1 \sigma $ error bars in
       the $y$ direction. The error bars in the $x$ direction are the bin sizes.
       The solid line shows the best fit King profile. The dashed vertical line
       shows the position of the core radius. The dot-dashed line shows the
   half-mass radius, and the dotted line shows the corresponding plummer profile.}
   \label{rdp}
   \end{center}
\end{figure}

To quantify the size and concentration of \obj/ we built the radial density
profile (RDP) by counting stars in annuli around its center. To increase
contrast relative to the background, only stars within 2$\sigma$ in
color away from the best fit isochrone were selected, where $\sigma$ is the
mean photometric error at a given magnitude. The data was also cut at $g < 24$
to avoid photometric incompleteness. The result is shown in Figure
\ref{rdp} where a very peaked distribution of stars is visible. 

In order to avoid the binning complications that arise when trying to fit a RDP
we chose a Maximum Likelihood (ML) approach. The method we use here follows
closely the one described in \citet{Martin}. 

The ML fit was carried out using stars in the range $0.0 \le R \le 12\arcmin$
and fitted both a King and a Plummer model. This limit was chosen to avoid the
large CCD gap of MegaCam. The best fitted models are shown as the solid and
dotted lines in Figure \ref{rdp}, respectively. For the King profile we derive
$r_{core} = 0.22^{+0.10}_{-0.06} \arcmin$ and $r_{tidal} =
7.85^{+4.15^*}_{-3.40}\arcmin$. And the Plummer profile fit yields $r_s =
0.60^{+0.16}_{-0.11} \arcmin$ and a total number of stars $M^* = 94.7 \pm 5.9$.
The uncertainty in each parameter is estimated using the likelihood-ratio
approximation, where likelihood values near the maximum approaches a chi-squared
distribution. The uncertainties are taken as the 90\% confidence level of the
chi-square distribution. For the Plummer profile we also kept the center of
\obj/ as a free parameter. This allowed us to find the best center and its
uncertainty. The best center was then used in the King profile fit. Notice that
the upper bound of the tidal radius uncertainty is outside the range of R
available in our observations. For simplicity we have adopted the upper bound as
the largest radii available, that is $12\arcmin$.

For the Plummer profile the half-mass radius is easily obtained from the
relation $r_h = 1.305r_s$, yielding the value of $r_h = 0.78^{+0.21}_{-0.14}
\arcmin$.  For the King model we made an estimate of the half-mass radius as
follows. We first integrated the profile from zero to the limiting radius and
subtracted the expected number of background stars. The result of this operation
yields the total number of observed stars that should belong to \obj/ $N_{obs}$.
We then compute the half-mass radius as the radius which contains $N_{obs}/2$
stars, again taking care to subtract off the expected background sources. We
obtain $r_h = 0.69^{+0.33}_{-0.25}\arcmin$. The two estimates of $r_h$ agree
within the uncertainties. Since the Plummer model is widely used in most
studies of satellites, we choose to use the half-mass radius estimated from it. 

\subsection{Total luminosity}

The total luminosity of \obj/ is dominated by bright red giant branch (RGB)
stars and possibly a few red clump (RC) stars, although Figure \ref{CMD_comp}
does not reveal any obvious RC star in the inner 150$\arcsec$ region.  The small
number of evolved stars in the RGB and, most especially, in the RC, indicates
that \obj/ is a low mass stellar system, more consistent with being a star
cluster than a dwarf galaxy. Given its low stellar mass, to properly estimate
the total luminosity of \obj/ one must take into account the background (and
foreground) star contamination.

We first built the observed luminosity function (LF) of \obj/ by counting stars
as a function of magnitude inside a circle of $150\arcsec$ radius from the
cluster center and in a ring with inner border at $300\arcsec$ and outer at
$550\arcsec$. We subtracted the area weighted star counts in the ring from the
area weighted counts in the inner circle. This results in an observed LF for
\obj/ already decontaminated from field stars. The adopted ring is 10 times
larger than the cluster circle to minimize fluctuations in the subtracted field
contamination. To build the LFs we again applied a magnitude cut of 
$g < 24.0$ to avoid incompleteness.

The total magnitude is obtained by direct integration of the background
decontaminated LF. Using the uncertainty on each bin of the LF and a bootstrap
method we derive the uncertainty on the total magnitude. The final value is $M_V
= -1.21 \pm 0.66$ for \obj/. Coupling the $M_V$ value with the $r_h$ estimate
yields a surface brightness of $\mu_V = 25.90 $ mag/arcsec$^2$.  The value of
$M_V$ of \obj/ is comparable to the value obtained for Mu{\~n}oz 1.  This
luminosity is also comparable to that of SEGUE 3 \citep{belokurov10, fadely}.

\section{Summary and Discussion}

\begin{deluxetable}{r|rl}
\label{tab1}
\tablecolumns{3}
\tablewidth{0pc}
\tablecaption{Summary of the derived parameters for \obj/.}
\tablenum{1}
\tablehead{Parameter & Value & Unity}
\startdata
$\alpha_{J2000}$ & $22:10:43.15 \pm 0.3$ & h:m:s \\
$\delta_{J2000}$ & $14:56:58.8 \pm 2.0$ & $^\circ:\arcmin:\arcsec$ \\
$d_{\odot}$      & $31.9^{+1.0}_{-1.6}$ & kpc \\
$M_V$            & $-1.21\pm0.66$ & mag \\
$log(age/yr)$    & $10.07^{+0.05}_{-0.03}$ & dex \\
$[Fe/H]$         & $-1.58^{+0.08}_{-0.13}$ & dex \\
\hline
\sidehead{Plummer}
\hline
$r_s$            & $0.6^{+0.16}_{-0.11}$  & arcmin\\ 
$\Sigma_{bg}$    & $1.38^{+0.03}_{-0.05}$ & stars/arcmin$^2$\\
$M^*$            & $94.7 \pm 5.9$ & \\
$r_h$            & $0.78^{+0.21}_{-0.14}$ & arcmin \\  
\hline
\sidehead{King}
\hline
$r_c$            & $0.22^{+0.10}_{-0.06}$ & arcmin\\ 
$r_t$            & $7.85^{+4.15}_{-3.40}$ & arcmin\\ 
$\Sigma_{bg}$    & $1.32^{+0.09}_{-0.07}$ & stars/arcmin$^2$\\
$\Sigma_{c}$     & $162.76\pm5.0$         & stars/arcmin$^2$\\
$r_h$            & $0.69^{+0.33}_{-0.25}$ & arcmin  
\enddata
\end{deluxetable}

In this paper we report on the discovery of a new stellar system in the MW halo,
found the SDSS-III/BOSS footprint in the Southern Galactic hemisphere. Its
confirmation as a genuine stellar system required deep follow-up imaging from
CFHT. By means of a theoretical isochrone fit, we derived a heliocentric
distance of $31.9^{+1.0}_{-1.6}$ kpc , an old age of $11.7^{+1.4}_{-0.8}$ Gyr,
and a metallicity of [Fe/H] $= -1.58^{+0.08}_{-0.13}$. We also found that a King
profile provides a good description of its structure; the best fit profile has a
core radius of $r_c = 2.04^{+0.93}_{-0.56}$ pc$^1$, a limiting radius of $r_t =
72.84^{+38.51}_{-31.55}$ pc, and a projected half-mass radius of $r_h =
7.24^{+1.94}_{-1.29}$ pc obtained using a Plummer profile. We carefully
estimate the object total luminosity by means of direct integration of the
background decontaminated LF. With the aid of a statistical bootstrapping we
estimate the uncertainties on the total absolute magnitude of \obj/ leading to
the final value of $M_V = -1.21 \pm 0.66$. In Table 1 we present a summary of
the parameter derived for \obj/.
\footnotetext[1]{The scale transformation from sky to physical values is 9.28 pc/arcmin}

The total number of stars and absolute magnitude of \obj/ suggest that it is a
star cluster. Its size is larger than that of most clusters in the Galactic
system, either open or globular, or in M31 \citep{schilbach06, vanden10,
vanden11}. However, it falls close to the median radius when compared to outer
halo clusters in the Galaxy.  Its size is more typical of the diffuse star
clusters found by \citet{peng06} in early-type galaxies, but again with several
orders of magnitude difference in terms of luminosity.  It is an extremely low
luminosity cluster; In fact, \obj/ seems to be one of the faintest and lowest
surface brightness old stellar systems found so far in the MW. Only Munoz 1
has a lower surface brightness \citep{munoz}, and only Munoz 1 and Segue 3 have
lower absolute magnitudes.  Its location in the luminosity vs. size diagram
places \obj/ close to other systems identified as low luminosity outer halo
clusters, including Koposov 1 and AM4 \citep{fadely}.  

The stellar cloud that lies closest to \obj/ is the Hercules-Aquila
\citep{belokurov07}. The cloud is located at $l\ \sim\ 50^\circ$ although its
extension is poorly known, specially in the southern galactic hemisphere. However
the heliocentric distance of the cloud is $10-20$ kpc, making it unlikely that
\obj/ is associated with this halo structure. 

The small size of the object, and the lack of any clear evidence for
complex stellar populations make it unlikely that it is a dwarf galaxy
\citep{willman12}.  We note that this locus is at the confluence of the branches
filled by classical globular clusters and MW dwarfs, as shown for instance in
\cite{maconnachie}.  Spectra from individual stars are being obtained and will
allow measurement of metallicity spread and line-of-sight velocities, which may
help determine the dynamical mass of \obj/ and constrain its stellar population.

\acknowledgments

Based on observations obtained with MegaPrime/MegaCam, a joint project of CFHT
and CEA/DAPNIA, at the Canada-France-Hawaii Telescope (CFHT) which is operated
by the National Research Council (NRC) of Canada, the Institute National des
Sciences de l'Univers of the Centre National de la Recherche Scientifique of
France, and the University of Hawaii.

LNdC acknowledges the support of FINEp grant 01.09.0298.00
0351/09, FAPERJ grants E-26/102.358/2009, E-26/110.564/2010, and
E-26/111.786/2011 and CNPq grants 304.202/2008-8 and 400.006/2011-1.

Funding for SDSS-III has been provided by the Alfred P. Sloan Foundation, the
Participating Institutions, the National Science Foundation, and the U.S.
Department of Energy Office of Science. The SDSS-III web site is
http://www.sdss3.org/.  SDSS-III is managed by the Astrophysical Research Consortium for the
Participating Institutions of the SDSS-III Collaboration including the
University of Arizona, the Brazilian Participation Group, Brookhaven National
Laboratory, University of Cambridge, Carnegie Mellon University, University of
Florida, the French Participation Group, the German Participation Group, Harvard
University, the Instituto de Astrofisica de Canarias, the Michigan State/Notre
Dame/JINA Participation Group, Johns Hopkins University, Lawrence Berkeley
National Laboratory, Max Planck Institute for Astrophysics, Max Planck Institute
for Extraterrestrial Physics, New Mexico State University, New York University,
Ohio State University, Pennsylvania State University, University of Portsmouth,
Princeton University, the Spanish Participation Group, University of Tokyo,
University of Utah, Vanderbilt University, University of Virginia, University of
Washington, and Yale University. 

{}

\end{document}